\documentclass[11pt]{article}
\usepackage{hyperref}

\usepackage{amssymb,amsmath} 
\usepackage{graphicx}
\usepackage{subfigure}

\usepackage{color}

\newcommand{\beq}[1]{  \begin{equation} \label{#1} }  
\newcommand{\eeq}{     \end{equation}}  
\newcommand{\bal}[1]{\begin{align} \label{#1} }

\newtheorem{thm}{Theorem}
\newtheorem{lem}{Lemma}
\newcommand{\rf}[1]{(\ref{#1})}

\def\bd#1{\mbox{\boldmath$\displaystyle\mathbf{#1}$} }
\def\div{\operatorname{div}} 
\def\Div{\operatorname{Div}} 
\def\dd{\operatorname{d}} 
\def\tr{\operatorname{tr}} 
\def\eps{\epsilon} 
\def\whbf#1{\widehat{\bf #1}}
\def\g{g}

\renewcommand{\appendix}{
  \setcounter{section}{0}\renewcommand{\thesection}{\Alph{section}}
  \section*{Appendix} }

\def\Appendix#1{
  \setcounter{equation}{0}
  \renewcommand{\theequation}{\thesection.\arabic{equation}}
  \section{#1}  }

\begin{document} 
\def\singlespacing{\baselineskip=13pt}	\def\doublespacing{\baselineskip=18pt}
\singlespacing

\title{Acoustic cloaking theory}  

\author{ Andrew N. Norris 
 \\ 
 Mechanical and Aerospace Engineering, \\
	Rutgers University, Piscataway NJ 08854-8058, USA \,\, norris@rutgers.edu}


\maketitle

\begin{abstract}

An acoustic cloak envelopes an object so that sound incident from all directions  passes through and around the cloak as though the object were not present.   
A theory of acoustic cloaking is developed using the transformation or change-of-variables method for mapping the cloaked region to a point with vanishing scattering strength.  
We show that the acoustical parameters in the cloak  must be anisotropic: either the mass density or the  mechanical stiffness or  both.  If the stiffness is isotropic, corresponding to a fluid with a single bulk modulus, then the inertial density must be infinite at the inner surface of the cloak.  This requires an infinitely massive cloak.    We show that  perfect cloaking can be achieved with finite mass  through the use of anisotropic stiffness.  The generic class of anisotropic material required is known as a pentamode material.  If the transformation deformation gradient is symmetric then the pentamode material parameters are explicit, otherwise its properties depend on a stress like tensor which  satisfies a static equilibrium equation.   For a given transformation mapping the  material composition of the cloak is not uniquely defined, but the phase and wave speeds of the pseudo-acoustic waves in the cloak  are unique.    Examples are given from  2D and 3D. 

\end{abstract}

\section{Introduction}

The observation that the electromagnetic equations remain invariant under spatial transformations 
is not new.  Ward and Pendry \cite{Ward96} used it for numerical purposes, but the result was known to Post \cite{Post62} who  discusses it in his book, and it was probably known far earlier.  The recent interest in passive cloaking and invisibility is due to the fundamental result of 
Greenleaf, Lassas an Uhlmann \cite{Greenleaf03,Greenleaf03b} that singular transformations 
could lead to  cloaking for conductivity.  Not long after this important discovery, 
Pendry et al. \cite{Pendry06} and Leonhardt \cite{Leonhardt06} made the key observation that  
singular transformations could be used to achieve cloaking of electromagnetic waves. 
These results and others have generated significant interest in the possibility of passive  acoustic cloaking.  

\begin{figure*}[htbp]
\centering
\includegraphics[width=5.0in , height=1.8in 					]{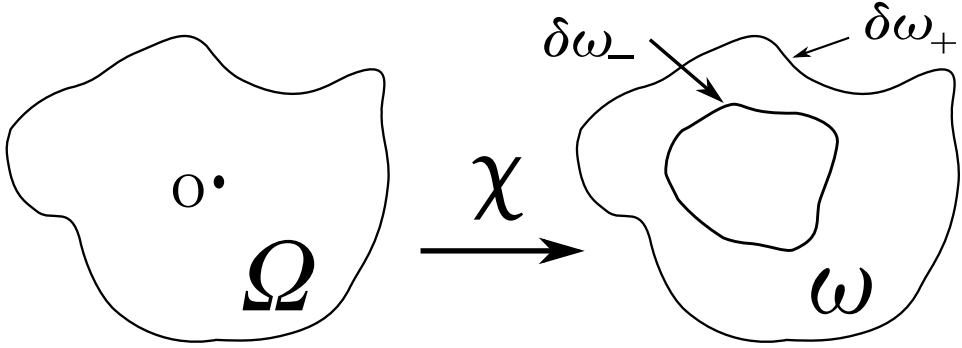} 
	\caption{The undeformed simply connected region $\Omega$ is transformed by the mapping ${\bd \chi}$ into the multiply connected cloak  $\omega$. Essentially, a single point ${\bd O}$ is transformed into a hole (the invisible region) surrounded by the cloak $\omega$.  The outer boundary $\partial \omega_+$ is coincident with $\partial \Omega_+ (= \partial \Omega)$ and the inner boundary $\partial \omega_-$ is the image of the  point ${\bd O}$.  Apart from ${\bd O}$ and $\partial \omega_-$ the mapping is everywhere one-to-one and differentiable. 	} 
		\label{f1}  
\end{figure*}

Acoustic cloaking is considered here in the context of the so-called transformation or change-of-variables method.   The transformation  deforms  a region  in such a way that the mapping is one-to-one everywhere except at a single point, which is mapped into the cloak inner boundary, see Figure \ref{f1}.
The acoustic problem is for the  infinitesimal  pressure $p({\bd x},t)$ which satisfies the  scalar   wave equation in the surrounding fluid,
\beq{91}
\nabla^2 p - \ddot{p} = 0. 
\eeq
The basic idea is to alter the cloak's acoustic properties (density, modulus) so that the modified  wave equation in $\omega$  mimics the exterior equation \rf{91} in the \emph{entire} region  $\Omega$.    This is achieved if the  spatial mapping of the simply connected region  $\Omega $  to the multiply connected cloak 
$\omega$  has the property  that the modified equation in $\omega$ when expressed in $\Omega$ coordinates has exactly the form of \rf{91} at every point in $\Omega$.  

The objective here is to answer the question: What type of material is required to realize these unusual properties that make  an acoustic cloak?    While cloaking cannot occur if the bulk modulus  and density are simultaneously  scalar quantities (see below), it is possible to obtain acoustical cloaks by assuming the mass density is anisotropic \cite{Cummer07,Chen07,Cummer08}.   
A  tensorial density is not ruled out on fundamental  grounds  \cite{Milton06} and in fact there is a strong physical basis for anisotropic inertia.  For instance, Schoenberg and Sen \cite{Schoenberg83} showed that the  inertia tensor in a medium comprising alternating fluid constituents is transversely isotropic with elements $\langle \rho \rangle$ in the  direction normal to the layering, and $\langle \rho^{-1} \rangle^{-1}$ in the transverse direction, where  $\langle \cdot \rangle$ is the spatial average.   Anisotropic effective density can arise from other microstructures, as discussed by Mei et al. \cite{Sheng07} and by Torrent  and S\'{a}nchez-Dehesa \cite{Torrent08}. 
The general context for anisotropic inertia is the   Willis equations of elastodynamics \cite{Milton07} which  Milton et al. \cite{Milton06} showed  are the natural counterparts of  the EM equations that  remain invariant  under spatial transformation.   Acoustic cloaking has been demonstrated, theoretically at least, in both 2D and 3D:   a spherically symmetric   cloak was discussed  by Chen and Chan \cite{Chen07} and by Cummer et al. \cite{Cummer08}, while   Cummer and Schurig \cite{Cummer07} described a 2D cylindrically symmetric acoustic cloak.  These papers use a  linear transformation  based on prior EM results    in 2D \cite{Schurig06}.  

Cloaks based on  anisotropic density in combination with the inviscid acoustic pressure constitutive relation (bulk modulus) will be called  Inertial Cloaks (IC). 
The fundamental  mathematical identity behind the IC is the observation of 
Greenleaf et al. \cite{Greenleaf07}  
 that the scalar wave equation is mapped into the following form in the deformed cloak region
\beq{92}
\frac1{\sqrt{|g|}} \frac{\partial ~}{\partial x_i}
\bigg( \sqrt{|g|} \, g^{ij} \frac{\partial p}{\partial x_j} \bigg) - \ddot{p} = 0,
\quad {\bd x} \in \omega .
\eeq
Here $g = (g_{ij})$ is the Riemannian metric with $|g|= \det (g_{ij})$, $(g^{ij}) = 
 (g_{ij})^{-1}$.   The reader familiar with differential geometry will recognize the first term in eq. \rf{92} as the Laplacian in curvilinear coordinates. 
Comparison of the transformed wave equation \rf{92} with the IC wave  equation provides explicit expressions for the IC density tensor and the bulk modulus \cite{Greenleaf08}.

  We will derive an identity equivalent to \rf{92} in Section \ref{sec1} using an alternative formulation  adapted from the theory of finite elasticity.   A close examination of the anisotropic density of the IC shows that its volumetric  integral, the total mass, must be infinite for perfect cloaking.   This raises grave questions about the utility of the IC.  The remainder of the paper provides a solution to this quandary.  The main result is that the IC is a special case of a more general class of acoustic cloaks, defined by anisotropic inertia combined with anisotropic stiffness.   The latter is obtained through the use of pentamode materials (PM) \cite{Milton95}. In the same way that an ideal acoustic fluid can be defined as the limit of an isotropic elastic solid as the shear modulus tends to zero, there is a class of limiting anisotropic solids with five (hence penta) easy modes of deformation analogous to shear, and one non-trivial mode of stress and strain.  The general cloak comprising PM and IC is called the PM-IC model.   The additional degrees of freedom provided by the PM-IC allow us to avoid the infinite mass dilemma of the IC. 
  
We begin in Section \ref{sec1} with a new derivation  of the inertial cloak (IC) model, and a discussion of the infinite mass dilemma.   Pentamode materials are introduced in Section \ref{sec2} where it is shown that they display simple wave properties, such as an ellipsoidal slowness surface.  The intimate connection between PM and acoustic cloaking follows from Theorem \ref{thm1} in Section \ref{sec3}.    Properties of the generalized PM-IC model for cloaking are developed in Section \ref{sec3} through the use of an example cloak that can be pure IC or pure PM as a parameter is varied.  Further examples are given in Section \ref{sec4}, with a concluding summary of the generalized acoustic cloaking theory in Section \ref{sec5}. 



\section{The inertial cloak}\label{sec1}

The transformation from $\Omega$ to $\omega$ is described by the point-wise deformation 
from  ${\bd X}\in \Omega$ to   ${\bd x} = {\bd \chi}({\bd X})\in \omega$.   In the language of finite elasticity, ${\bd X}$ describes a particle position  in the Lagrangian or undeformed configuration, and ${\bd x}$ is particle location in the Eulerian or   deformed  physical state.  The transformation or mapping defined by ${\bd \chi}$ is one-to-one and invertible except at the single point ${\bd X}={\bd O}$, see Figure \ref{f1}. 
 We use $\nabla$, $\nabla_X$ and $\div$, $\Div$ to indicate the gradient and divergence operators in ${\bd x}$  and ${\bd X}$, respectively.  The component form of $\div {\bf A}$ is $\partial A_i /\partial x_i$ or 
 $\partial A_{ij} /\partial x_i$ when  ${\bf A}$ is a vector and a second order tensor-like quantity, respectively. 
The deformation gradient is defined 
${\bd F} = \nabla_X {\bd x}$  with inverse ${\bd F}^{-1} = \nabla {\bd X}$, 
or in component form $F_{iI}  = \partial x_i /\partial X_I$, $F_{Ii}^{-1}  = \partial X_I /\partial x_i$.    
The Jacobian of the deformation is $
J = \det {\bd F} = |{\bd F}|$,  or in terms of volume elements in the two configurations, $J = \dd v/\dd V$. The  polar decomposition implies ${\bd F} = {\bd V} {\bd R} $, where ${\bd R}$ is proper  orthogonal ($ {\bd R}{\bd R}^t = {\bd R}^t {\bd R} = {\bd I}$, $\det {\bd R}= 1$) and the left stretch tensor ${\bd V}\in$Sym$^+$ is the  positive definite solution of ${\bd V}^2 = {\bd F}{\bd F}^t$.  The analysis is as far as possible independent of the spatial dimension $d$, although applications are restricted to $d=2$ or $d=3$.

The principal result for the inertial cloak is 
\begin{lem}\label{lem1}
\beq{+32}
\nabla_X^2 p = J \, \div \big( J^{-1} {\bd V}^2 \nabla p\big). 
\eeq
\end{lem}
Proof: 
The right hand side can be expressed 
\beq{104}
J \, \div \big( J^{-1} {\bd F}{\bd F}^t  \nabla p\big)
=   JJ^{-1} ({\bd F}^t \nabla ) \cdot \big( {\bd F}^t  \nabla p\big) + 
J ({\bd F}^t  \nabla p)\cdot \div \big(  J^{-1}{\bd F} \big). 
\eeq
Using the chain rule in the form 
${\bd F}^t \nabla  = \nabla_X$ or  $\nabla  = {\bd F}^{-t}\nabla_X  $ implies that 
${\bd F}^t \div \big( {\bd F}^t  \nabla p\big) = \Div \nabla_X p$ which is 
$\nabla_X^2 p$.  The proof follows from the 
identity (see Problems 2.2.1 and 2.2.3 in \cite{Ogden84})
\beq{+21}
\div \big( J^{-1} {\bd F}\big) = 0.  
\eeq

\subsection{Cloak acoustic parameters}
 
The connection with acoustics is made by identifying the field variable $p$ in Lemma \ref{lem1} as the  acoustic pressure.  The  cloak comprises an inviscid  fluid with 
bulk modulus $K({\bd x})$ such that the pressure satisfies the standard relation 
\beq{931}
\dot{p} = - K \div {\bd v}, 
\eeq
where ${\bd v}({\bd x},t)$ is  particle velocity.   The inertial cloak is defined by the 
assumption that the momentum balance involves a symmetric second order   inertia tensor  ${\bd \rho}$ according to 
\beq{932}
{\bd \rho} \dot{\bd v} = - \nabla p , 
\eeq
Although this is a significant departure from classical acoustical theory in assuming an anisotropic mass density,  it is by no means  unprecedented.    Based on the analysis of  Schoenberg and Sen \cite{Schoenberg83}, 
a spatially varying tensor ${\bd \rho}$ could possibly be achieved by small pockets of layered fluid  separated by   massless impermeable membranes.

Eliminating the velocity between eqs. \rf{931} and \rf{932} gives a single equation for the pressure 
\beq{933}
K \div \big( {\bd \rho}^{-1}\nabla p\big) - 
  \ddot{ p} = 0 , \qquad
  {\bd x} \in  \omega. 
\eeq
Consider the uniform wave equation in $\Omega$: 
\beq{+44}
\nabla_X^2 p - \ddot{ p} = 0,  \qquad
  {\bd X} \in  \Omega. 
\eeq
   Using Lemma \ref{lem1} we can express this in the  deformed physical description as eq. \rf{933} where
the bulk modulus and inertia tensor are  
\beq{+4}
K = J, \qquad
   {\bd \rho} = J  {\bd V}^{-2}.  
\eeq
For a given deformation ${\bd F}$, the identities \rf{+4} define the unique cloak with spatially varying material parameters $K$  and ${\bd \rho}$ each defined by the deformation gradient.  
We note the following identity which  is independent of  ${\bd F}$: 
\beq{+2}
  \det {\bd \rho} =K^{d-2}.  
\eeq

Could the cloak possibly have isotropic density?  That is, could the cloak be described by a standard acoustic fluid  with two scalar parameters, density and bulk modulus?  The identity $ {\bd \rho} = J  {\bd V}^{-2}$ means that $ {\bd \rho} = \rho {\bd I}$ can occur only if ${\bd V}$ is a multiple of the identity, 
  ${\bd V}= w {\bd I}$ for some scalar $w=w({\bd x})$.   The deformation of $\Omega$ into  the smaller region $\omega$ could  certainly be accomplished at {\it some  but not all points} by this  deformation, which corresponds to a uniform  contraction or expansion, with rotation.   However, the deformation near the inner surface of the cloak cannot be of this form.  In fact, the deformation in  the neighbourhood of ${\bd X}={\bd O}$ must be extremely nonuniform and anisotropic.  We will discuss this below when we examine a fundamental and severe deficiency  of the inertial cloak model. 
  
\subsection{Continuity between the cloak and the acoustic fluid}

 Let $\dd s$, 
${\bd n}$ and $\dd S $, ${\bd N}$ denote the area element and unit normal to the outer boundary   $ \partial \omega_+$, $  \partial \Omega_+ (=\partial \omega_+)$, respectively.  These are related by  the deformation through Nanson's formula
\cite{Ogden84} 
$
{\bd N}  \dd S  = J^{-1} {\bd F}^{t}{\bd n}  \dd s  $. 
The nature of the cloak requires that the outer surface is identical in either description, since both must match with the exterior fluid.  We therefore require that $\dd s = \dd S $ at every point on the outer surface, or
\beq{902}
{\bd N}    = J^{-1} {\bd F}^{t}{\bd n} \quad \text{on } \partial \omega_+ = \partial \Omega_+ ,
\eeq
and  eq. \rf{+4} then implies 
\beq{972}
     {\bd \rho}^{-1} {\bd n} = {\bd F}{\bd N} \quad \text{on } \partial \omega_+ = \partial \Omega_+.
\eeq
Equation \rf{972} is a purely kinematic condition.

The interior of the cloak mimics the  wave equation in the exterior fluid. The final requirement that the cloak will be acoustically ``invisible" is that the pressure and normal velocity match across the outer surface separating the fluid and cloak.  These two continuity conditions arise from the balance of force (normal traction) per unit area and the constraint of particle continuity. The  condition for pressure is  simply that $p$ is continuous across the outer surface, whether one uses the wave equation in physical space, \rf{933}, or its counterpart in the undeformed simply connect region, \rf{+44}. 
As for the kinematic condition, consider its equivalent, the  continuity of normal acceleration.  This is  $\dot{v}_n
= {\bd n}\cdot \dot{\bd v}$ in physical space, and using eq. \rf{932} it becomes $\dot{v}_n = - {\bd n}\cdot {\bd \rho}^{-1} \nabla p$, 
which must match with  $- {\bd n}\cdot \nabla p$ in the fluid.  Alternatively, 
eq.  \rf{972}  and the relation ${\bd F}^t \nabla  = \nabla_X$  imply, as expected,  
\beq{808}
\dot{v}_n = - {\bd n}\cdot {\bd \rho}^{-1} \nabla p
=  -  {\bd N} {\bd F}^t  \cdot \nabla p
=  -   {\bd N}  \cdot \nabla_X p .
\eeq
The final term is simply the normal acceleration in the undeformed description. 

In summary, the  continuity  conditions at the outer surface in the physical description are 
\beq{538}
\big[ p \big] = 0, \qquad 
\big[ {\bd n}\cdot {\bd \rho}^{-1} \nabla p \big]=0
\quad \text{on } \partial \omega_+ .  
\eeq

  
\subsection{Example: a rotationally symmetric inertial cloak }
Consider the inverse deformation 
\beq{551}
{\bd X} = f(r)\hat{\bd x},
\eeq
where $\hat{\bd x} = {\bd x}/r$ and $r = |{\bd x}|$. Using ${\bd F}^{-1} = \nabla {\bd X}$ implies
\beq{552}
{\bd F} =  (1/f') {\bd I}_r +  (r/f){\bd I}_\perp,
\eeq
where $f' = \dd f/\dd r$ 
and the second order tensors are 
${\bd I}_r =  \hat{\bd x} \otimes \hat{\bd x}$, ${\bd I}_\perp = {\bd I}- \hat{\bd x} \otimes \hat{\bd x}$.  The bulk modulus and mass density in the cloak  follow from eq. \rf{+4} as 
\beq{553}
K = \frac1{f'} \big( \frac{r}{f}\big)^{d-1} , \qquad
   {\bd \rho} = \big( \frac{r}{f}\big)^{d-1} 
   \big(  f' {\bd I}_r + \frac{f^2}{r^2f'}{\bd I}_\perp \big).
\eeq
The anisotropic inertia has the form  
\beq{555}
{\bd \rho} = \rho_r {\bd I}_r + \rho_\perp{\bd I}_\perp, 
\eeq
where the radial and azimuthal  principal values $\rho_r$ and $\rho_\perp$ can be read off from eq. \rf{553} as functions of $f$. 

Introducing the radial and azimuthal phase speeds: 
$c_r  =  \sqrt{  {K}/{\rho_r} }$ and 
 $c_\perp  =  \sqrt{ {K}/{\rho_\perp} }$,  
the mass density tensor can then be expressed  
$
{\bd \rho} =  K  
\big(  c_r^{-2} {\bd I}_r  +  c_\perp^{-2}{\bd I}_\perp \big) $. 
The quantity $K \rho_r $ is the square of the radial acoustic impedance, $z_r \equiv \sqrt{K \rho_r}$.  
Equation  \rf{+2} implies that the identity $z_r = c_\perp^{d-1}$  is required for cloaking. 
 The three equations   \rf{553} for $K$, $\rho_r$ and $\rho_\perp$ in terms of $f$ can be replaced by
 the universal relation \rf{+2}, i.e., 
 \beq{111}
 \rho_r \rho_\perp^{d-1} = K^{d-2}, 
 \eeq
  along with simple expressions for the wave speeds in terms of $f$: 
\beq{110}
c_r   =    \frac{1}{f'} , 
 \qquad
c_\perp =      \frac{r}{f}  . 
\eeq
We will see later that the phase and the wave (group velocity)  speeds in the principal directions are  identical.  
Note that $f'$ is required to be positive. 
 The original quantities can be expressed in terms of the phase speeds as 
 \beq{-3}
   \rho_r =  c_r^{-1}{c_\perp^{d-1}},\qquad
   \rho_\perp =c_r c_\perp^{d-3}   ,
   \qquad
   K = c_r c_\perp^{d-1} .
 \eeq
One  could, for instance, eliminate $f$ as the fundamental variable defining the cloak in favor of  $c_\perp (r)$, from which all other quantities can be determined from the differential equation  relating the speeds: 
$
\big(  {r}/{  c_\perp}\big)' = 1/c_r $.

We assume the cloak occupies $\omega = \{{\bd x}: 0< a\le |{\bd x}| \le b\}$ with uniform  acoustical  properties   $K=1$, ${\bd \rho} = {\bd I}$ in the exterior.  
The  areal matching condition    \rf{972} with  ${\bd n} = {\bd N} = \hat{\bd x}$ is satisfied by 
${\bd F}$ and ${\bd \rho}$ of eqs. \rf{552} and  \rf{553} if   $f$ is continuous across the boundary, which is accomplished by requiring $f(b) = b$.  The pressure and velocity continuity conditions 
\rf{538} become
\beq{539}
\big[ p \big] = 0, \qquad 
\big[ \frac1{f'} \frac{\partial p}{\partial r} \big]=0
\quad \text{on } r=b.  
\eeq
 
Note that the cloak  density is  isotropic  if $c_r = c_\perp$, 
 which requires that $f' =   f/r$.  Thus
$f = \gamma r$ with  
  $\gamma$ constant, but the   outer boundary condition $f(b)=b$ implies  $\gamma  = 1$, which is the trivial undeformed configuration. 

Perfect cloaking requires that $f$ vanish at $r= a$.  It is clear that $c_\perp$   blows up as 
$r\downarrow a$, as does the product $K \rho_r$.  In order to examine the individual behaviour of $K $ and $\rho_r$ consider $f \propto (r-a)^\alpha $ near $a$ for  $\alpha $ constant and non-negative.  No value  of $\alpha >0$ will keep the radial density $\rho_r$ bounded, although  the unique choice $\alpha = 1/d$ ensures that the  bulk modulus $K (a)$ remains finite and non-zero.    
Note that the azimuthal density $\rho_\perp$ has a finite limit in 2D for power law decay 
$f \propto (r-a)^\alpha $, while $\rho_\perp$ remains finite in 3D  iff $\alpha\le 1$, otherwise it blows up.   Similarly, the radial phase speed scales as $c_r \propto (r-a)^{1-\alpha} $, which remains finite  for $\alpha\le 1$,  blowing up otherwise.
These results are summarized in Table 1. 


\begin{table}		\label{tab1}
\begin{center}	
\caption{Behaviour of quantities near the inner surface $r=a$ for the scaling
$f \propto \xi^\alpha $ as $\xi= r-a \downarrow 0$.  The total radial mass $m_r$ is defined in eq. \rf{43}. }
\begin{tabular}{ccccccc} \hline  
 \, dim \,  & $\rho_r$ & $\rho_\perp$ &  $c_r$ & $c_\perp$ & $K$ & $m_r$ 
  \\ \hline 
 2 &   $\xi^{-1}$ & $\xi$ & $\xi^{1-\alpha}$ & $\xi^{-\alpha}$ & $\xi^{1-2\alpha}$ & $\ln \xi$ 
 \\
 3 &  \, $\xi^{-1-\alpha}$\, &\, $\xi^{1-\alpha}$\, & \,$\xi^{1-\alpha}$ \,&\, $\xi^{-\alpha}$ \,&\, $\xi^{1-3\alpha}$\,  &\, $\xi^{-\alpha}$ \,
 \\
\hline 		 
\end{tabular}
\end{center}
\end{table}

We use a nondimensional measure of the total mass in the cloak: 
$
{\bd m} \equiv   (\text{vol}(\omega))^{-1} \int_\omega \dd v {\bd \rho} $.
The total mass is isotropic for the symmetric deformation and configuration considered here: 
$ {\bd m} = \frac1{d}(m_r + (d-1)m_\perp) {\bd I}$ where 
$ (m_r, m_\perp) = (\text{vol}(\omega))^{-1} \int_\omega \dd v  (\rho_r, \rho_\perp) $. 
Assuming for the moment that $f(a)$ is nonzero, i.e., a near-cloak \cite{Kohn08}, then 
\beq{43}
m_r 
=
\begin{cases} 
\frac{2}{b^2-a^2}\big[ 
b^2 \ln f(b) - a^2 \ln f(a) - 2 \int_a^b \dd r \, r  \ln f(r)
\big] 
, & 2D,
\\
\frac{3}{b^3-a^3}\big[ 
\frac{a^4}{f(a) } - \frac{b^4}{f(b) } + 4\int_a^b \dd r \, \frac{r^3 }{f(r)}
\big]  ,  & 3D. 
\end{cases}
\eeq
These forms indicate not only that $m_r \rightarrow \infty$ as $f(a) \rightarrow 0$, but also the form of the blow-up.  To leading order, 
$m_r = \frac{2a^2}{b^2-a^2}\ln \frac{1}{f(a)} + \ldots$ in 2D and 
$m_r = \frac{3a^3}{b^3-a^3} \frac{a}{f(a)} + \ldots$ in 3D.  The blow-up of $m_r$ occurs no matter how $f$ tends to zero.  The infinite mass is an unavoidable  singularity.

\subsection{A massive problem with inertial cloaking}\label{1.4}

Table 1 and the example above illustrate a potentially grievous issue:  infinite mass is required for perfect cloaking in the IC model.  We now show that the  problem  is not specific to the   rotationally symmetric cloak but is common to all inertial cloaks. 
Consider a ball of radius $\eps$ around ${\bd X}= {\bd O}$.  Its volume $\dd V=$O$(\eps^d)$  is mapped to a volume with  inner surface defined by the finite cloak inner boundary $\partial \omega_-$  and outer surface a distance O$(\eps^\beta)$ further out, where $\beta >0$ is a local scaling parameter, assumed constant (in terms of the example above and Table 1, $\beta = 1/\alpha)$.  The mapped current volume is then $\dd v=$O$(\eps^\beta)$ so that  $J=\dd v/\dd V =$O$(\eps^{\beta-d})$.   The eigenvalues of  ${\bd V}$ are   
$\lambda _1=$O$( \eps^{\beta -1} )$, 
$\lambda _{2,\ldots, d}=$O$(\eps^{-1} )$. The bulk modulus and the principal
 values of the density matrix are  therefore
 \beq{012}
 K = \text{O}(\eps^{\beta-d}),
 \quad
 \rho_1=\text{O}(\eps^{2-d -\beta}),
 \quad
\rho _{2,\ldots, d}=\text{O}(\eps^{\beta-d +2} ).
\eeq
The principal value $\rho_1$ blows up whether $d=2$ or $d=3$.  Furthermore, the \emph{total mass} associated with $\rho_1$ in the mapped volume is $m_1 = $O$(\eps^{2-d} )$ which blows up in 3D, and a more careful analysis for 2D similar to that for the rotationally symmetric case shows $m_1 = $O$ (|\ln \eps | )$.

In summary, the inertial cloak theory, while consistent and formally sound, reveals an underlying and ``massive" problem. 
We will  show how this  can be circumvented by using a more general cloaking theory which allows for anisotropic stiffness (elasticity) in addition to, or instead of, the anisotropic inertia.   The  anisotropic elastic material required is of a special type, called a pentamode material \cite{Milton95}, which is introduced next.

\section{Pentamode materials}\label{sec2}

We consider  Hooke's law in 3D in the form  
$\hat{\bd \sigma} = \whbf{C} \hat{\bd \varepsilon} $, where  the 6-vectors of stress and strain, and the associated 6$\times$6 matrix of moduli are
\beq{002}
\hat{\bd \sigma} 
=  \begin{pmatrix}
\sigma_{11}
\\  
\sigma_{22}
\\  
\sigma_{33}
\\  
\sqrt{2} \sigma_{23}
\\  
\sqrt{2} \sigma_{31}
\\  
\sqrt{2} \sigma_{12}
\end{pmatrix}, 
\quad  
\hat{\bd \varepsilon} 
=  \begin{pmatrix}
\varepsilon_{11}
\\  
\varepsilon_{22}
\\  
\varepsilon_{33}
\\  
\sqrt{2} \varepsilon_{23}
\\  
\sqrt{2} \varepsilon_{31}
\\  
\sqrt{2} \varepsilon_{12}
\end{pmatrix}, 
\quad  
\whbf{C}   =  \begin{pmatrix}
C_{11} & C_{12} & C_{13} & 
 2^{\frac12} C_{14} & 2^{\frac12} C_{15} & 2^{\frac12} C_{16} 
\\ 
 & C_{22} & C_{23} & 
 2^{\frac12} C_{24} & 2^{\frac12} C_{25} & 2^{\frac12} C_{26} 
\\ 
 &  & C_{33} & 
 2^{\frac12} C_{34} & 2^{\frac12} C_{35} & 2^{\frac12} C_{36} 
\\ 
 &  &  & 2C_{44} & 2C_{45} & 2C_{46}
\\ 
 {\rm S} & {\rm Y} & {\rm M} & & 2C_{55} & 2C_{56}
\\ 
& & & & & 2C_{66}
\end{pmatrix} \, .  
\nonumber
\eeq
The $\sqrt{2}$ terms ensure that products and norms are preserved, e.g. 
$C_{ijkl}C_{ijkl} = \tr \whbf{C}^t\whbf{C}$.   

A pentamode material (PM)  is rank one, or in other words, five of the six eigenvalues of $\whbf{C}$ vanish \cite{Milton95}. The one remaining positive eigenvalue is therefore
\beq{003}
3 \widetilde{K} \equiv  \tr \whbf{C} = C_{ijij}= C_{11}+C_{22}+C_{33}+ 2 ( C_{44}+C_{55}+C_{66}). 
\eeq
Accordingly, the moduli can be defined by the stiffness $\widetilde{K}$ and a normalized 6-vector  $\hat{\bd s}$, 
\beq{004}
 \whbf{C} =  \widetilde{K}\,  \hat{\bd s}\hat{\bd s}^t,\qquad 3 = \hat{\bd s}^t\hat{\bd s} . 
\eeq
The stress is described by a single scalar,  $\hat{\bd \sigma} = \sigma \hat{\bd s}$ with $\sigma = \hat{K}  \varepsilon $
and $ \varepsilon = \hat{\bd s}^t \hat{\bd \varepsilon} $. 
Thus, 
\beq{0045}
{\bd C} = \widetilde{K} \, \widetilde{\bd S}\otimes \widetilde{\bd S}
\qquad
\Leftrightarrow
\qquad
C_{ijkl} = \widetilde{K} \,\widetilde{S}_{ij} \widetilde{S}_{kl},
\qquad
\widetilde{\bd S}
=  \begin{pmatrix}
\hat{s}_1 & \frac1{\sqrt{2}}\hat{s}_6 & \frac1{\sqrt{2}}\hat{s}_5 
\\
\frac1{\sqrt{2}}\hat{s}_6 & \hat{s}_2 & \frac1{\sqrt{2}}\hat{s}_4
\\
\frac1{\sqrt{2}}\hat{s}_5 & \frac1{\sqrt{2}}\hat{s}_4 & \hat{s}_3
\end{pmatrix}.
\eeq
The pentamode material \cite{Milton06} is so named because there are five easy ways to deform it, associated with the eigenvectors of the five zero  eigenvalues of the elasticity stiffness.  Pentamodes obviously include isotropic acoustic fluids, for which the only stress-strain eigenmode is a hydrostatic stress, or pure pressure, and the five easy modes are all pure shear.   Milton and Cherkaev \cite{Milton95} describe  how pentamode materials can be realized from specific microstructures.

\subsection{Example: An orthotropic PM}
An elastic  material with  orthotropic symmetry  has nine non-zero elements in general: the six 
$  C_{ij} = C_{ji}$, $i,j = 1,2,3$, plus    $C_{44}$, $C_{55}$ and $C_{66}$.  We set these last  three (shear) moduli to zero.  
The stress ${\bd \sigma}$ must then be diagonal in the cartesian coordinate system, implying $\hat{s}_4= \hat{s}_5 = \hat{s}_6 = 0$, and therefore 
\beq{007}
\widetilde{K} = \frac13(C_{11}+C_{22}+C_{33}),
\qquad
 \widetilde{\bd S}
=  \widetilde{K}^{-1/2}\big( 
C_{11}^{1/2} {\bd e}_1\otimes {\bd e}_1
+ C_{22}^{1/2} {\bd e}_2\otimes {\bd e}_2
+ C_{33}^{1/2} {\bd e}_3\otimes {\bd e}_3
\big), \nonumber
\eeq
with  the following relations holding: 
$C_{12} = C_{11}^{1/2}C_{22}^{1/2}$, $C_{23} = C_{22}^{1/2}C_{33}^{1/2}$,
$C_{13} = C_{11}^{1/2}C_{33}^{1/2}$. 

\subsection{Compatibility condition  for pentamode materials}

The notation $\widetilde{K}$ and $\widetilde{\bd S}$ is used to signify the fact that 
the $\widetilde{\bd S}$ tensors are normalized by $\tr \widetilde{\bd S}^2 =3$ and therefore
 $\widetilde{K}$ is given by eq. \rf{003}.  We will not follow this normalization in general, but write 
 \beq{450}
{\bd C} =  {K} \,  {\bd S}\otimes  {\bd S}
\qquad
\Leftrightarrow
\qquad
C_{ijkl} =  {K} \, {S}_{ij}  {S}_{kl}. 
\eeq
In other words, the products in \rf{450} are the important physical quantities, not $K$ and ${\bd S}$ individually.   
The stress in the PM is always proportional to the  tensor ${\bd S}$ and only one strain element is significant, ${\bd S} : {\bd \eps}$.  The rank deficiency of the moduli, which  is apparent from \rf{004} or \rf{450}, means that there is no  
inverse strain-stress relation for the elements of ${\bd \eps}$ in terms of the elements of ${\bd \sigma}$.   

Static equilibrium of a pentamode material  under an applied load leads to a constraint on the spatial variability of the PM stiffness.  
Consider an inhomogeneous pentamode material with smoothly varying  ${\bd C} ( {\bd x})
= K_0 ( {\bd x}){\bd S}_0( {\bd x})\otimes  {\bd S}_0( {\bd x})$.    Under an applied static load the strain will also be spatially inhomogeneous, but the only part of the strain  that is important is the component 
along the PM eigenvector.  With no loss in generality we may put 
${\bd \varepsilon} ( {\bd x}) = w ( {\bd x}){\bd S}_0 $ for some scalar function $w$.  The stress is then 
${\bd \sigma} = q{\bd S}_0 $ where $ q ( {\bd x}) = w K_0 \tr( {\bd S}_0^2) = 
3 \widetilde{K} w$.  
Let ${\bd S} = q{\bd S}_0 $, then the static equilibrium condition $\div {\bd \sigma}=0$ becomes 
$\div {\bd S}=0$. 
Finally, the PM stiffness is ${\bd C} =  {K}  {\bd S}\otimes  {\bd S}$
where $K ( {\bd x}) = K_0 /( 3 \widetilde{K} w)^2$.

\begin{lem}\label{lem2}
The fourth order stiffness of a smoothly varying pentamode material can always be expressed 
${\bd C}  =  {K}    {\bd S}\otimes  {\bd S}$, where $K ( {\bd x}) >0$ and ${\bd S} ( {\bd x}) \in$Sym satisfies the static  equilibrium condition
\beq{040}
\div {\bd S}=0.
\eeq
\end{lem}
This identity also arises in a completely different manner later when we consider transformed wave equations. We say that the PM is of {\it canonical form} when eq. \rf{040} applies.  The decomposition of Lemma \ref{lem2} is unique up to a multiplicative constant. Thus, if a static load is applied to a PM expressed in canonical form, then the stress and strain are 
    ${\bd \sigma} ({\bd x}) = c_0 {\bd S} $ and 
    ${\bd \varepsilon} ( {\bd x}) = (c_0 3\widetilde{K})^{-1} {\bd S}  $, respectively, for constant $c_0$.  
    
In summary, stability under static loading places a constraint on 
 the PM moduli, which will turn out to be useful when we return to the cloaking problem.  The constraint means that the moduli 
 can in general be expressed in  canonical form.

\subsection{Dynamic equations of motion in a  PM}

The equations for small amplitude disturbances in a  PM with anisotropic mass density are 
\bal{--1}
{\bd \sigma} & = K \tr ( {\bd S}{\bd \varepsilon})\, {\bd S}, 
\\
{\bd \rho} \dot{\bd v} &= \div {\bd \sigma} .
\label{1}
\end{align}
These are respectively, the specific form of Hooke's law for a PM and   the  momentum balance   incorporating the inertia tensor. 
In order to make the equations look similar to those for an acoustic fluid, we  identify  the ``pseudo-pressure'' 
$p$ with the negative single  stress: $p = - K \tr ( {\bd S}{\bd \varepsilon})$.  The stress tensor becomes
\beq{-21}
{\bd \sigma} = -p {\bd S} ,  
\eeq
and the linear constitutive relation  can be written 
\beq{-23}
\dot{p} = - K{\bd S}:\nabla {\bd v}.    
\eeq
Equations \rf{1} and \rf{-23} imply that the pseudo-pressure satisfies the generalized acoustic wave equation 
\beq{207}
 K {\bd S} :\nabla \big(  {\bd \rho}^{-1}\div (   {\bd S} p )    \big) -  \ddot{p} =0. 
\eeq
This   reduces to the   acoustic equation \rf{933} with anisotropic inertia and isotropic stiffness when ${\bd S} = {\bd I}$.  Finally, assuming that the PM is in canonical form, so that ${\bd S}$  satisfies the equilibrium condition  \rf{040}, we have 
\beq{202}
 K {\bd S} :\nabla \big(  {\bd \rho}^{-1}  {\bd S}\nabla p    \big) -  \ddot{p} =0. 
\eeq

\subsection{Wave motion in a  PM}

The wave properties of pentamode materials are of interest since we will show that they can be used to make the acoustic cloak.  
Consider plane wave solutions for  displacement of the form 
$
{\bd u}({\bd x},t) = {\bd q} e^{ik( {\bd n}\cdot {\bd x} - v t)}$, 
for $|{\bd n}|=1$ and constant ${\bd q}$, $k$ and $v$, and uniform PM properties.  
Non trivial solutions of the equations of motion \rf{--1}-\rf{1} must satisfy 
\beq{-125}
\big(  K\,  ({\bd S}{\bd n})\otimes ({\bd S}{\bd n}) - {\bd \rho} v^2 \big) {\bd q}= 0.  
\eeq
The acoustical or Christoffel \cite{Musgrave}  tensor $K  ({\bd S}{\bd n})\otimes ({\bd S}{\bd n})$ is rank one  and it follows that  of the three possible solutions for $v^2$, only one is not zero, the quasi-longitudinal solution
\beq{-104}
v^2 =  K\,  {\bd n}\cdot  {\bd S} {\bd \rho}^{-1}{\bd S}{\bd n} , 
\qquad
{\bd q} = {\bd \rho}^{-1}{\bd S}{\bd n} . 
\eeq
The slowness surface is therefore an ellipsoid. 
 Standard arguments for waves in anisotropic solids \cite{Musgrave} show that the energy flux velocity (or wave velocity or ray direction) is 
\beq{37}
{\bd c} = v^{-1} K\,    {\bd S} {\bd \rho}^{-1}{\bd S}{\bd n}. 
\eeq
Note that this is in the direction ${\bd S}{\bd q}$, and satisfies  
${\bd c}\cdot {\bd n} = v$, a well known relation for generally anisotropic solids with isotropic density. 

As an example, consider the orthotropic PM with a density tensor of the same symmetry and coincident principal axes.
Then 
\begin{subequations}
\bal{-14}
v^2 =&  c_1^2 n_1^2  + c_2^2n_2^2  + c_3^2 n_3^2 , 
\\
{\bd c} =& v^{-1}\big( c_1^2 n_1 {\bd e}_1+ c_2^2 n_2 {\bd e}_2  + c_3^2 n_3 {\bd e}_3  \big), 
\\
{\bd q} =&  \rho_1^{-1/2} c_1 n_1 {\bd e}_1 +\rho_2^{-1/2} c_2 n_2 {\bd e}_2 +  \rho_3^{-1/2} c_3 n_3 {\bd e}_3, 
\end{align}
\end{subequations}
where $c_1^2 = C_{11}/\rho_1$, $c_2^2 = C_{22}/\rho_2$,  $c_3^2 = C_{33}/\rho_3$, 
and   $\rho_1$, $\rho_2$, $\rho_3$ are the principal inertias. 

\section{The general acoustic cloaking theory}\label{sec3}

We now show that the inertial cloak (IC) is but a special case of a much more general type of acoustic cloak.   While the IC depends upon the anisotropic inertia, the general cloaking model can have both anisotropic inertia and stiffness.   The additional degree of freedom is obtained by replacing the  pressure field with the scalar stress of a PM.   The general cloaking model is called  PM-IC. 

\subsection{The fundamental identity}

\begin{lem}\label{lem3}
Let  ${\bd P} \in$Sym be non-singular and ${\bd F}$ is the deformation gradient for the mapping 
${\bd X} \rightarrow  {\bd x}$ with $J = \det {\bd F}$, ${\bd V}^2 = {\bd F}{\bd F}^t$.  Then 
\beq{+13}
 \nabla_X^2 p = 
 J  {\bd P} :   \nabla \big( J^{-1}{\bd P}^{-1} {\bd V}^2\nabla      p \big)  , 
\eeq
iff ${\bd P}$ satisfies 
 \beq{+14}
 \div  {\bd P}  = 0.
 \eeq
\end{lem}
The proof is given in  Appendix A.  This clearly generalizes Lemma \ref{lem1}, and in the context of pentamode materials it implies
  
\begin{thm}\label{thm1}
The pressure $p$  satisfies a uniform wave equation in $\Omega$.  Under the transformation 
$\Omega \rightarrow \omega $ with $J = \det {\bd F}$, ${\bd V}^2 = {\bd F}{\bd F}^t$, $p$ satisfies the equation for the pseudo-pressure   of a pentamode material with stiffness ${\bd C}$ and anisotropic inertia ${\bd \rho}$: 
\begin{subequations}
\beq{140}
\nabla^2_X  p - \ddot{p}=0\quad \text{in }\Omega
\qquad
\Leftrightarrow
\qquad
K  {\bd S} :   \nabla \big( {\bd \rho}^{-1} {\bd S} \nabla   p \big) 
- \ddot{p}=0\quad \text{in }\omega,
\eeq
where
\beq{+14a}
 K = J , \qquad
 {\bd C}  = K {\bd S} \otimes  {\bd S},
 \qquad
 {\bd \rho}  = J {\bd S} {\bd V}^{-2}   {\bd S},
 \eeq
 and ${\bd S}$ satisfies 
 \beq{+14b}
 \div  {\bd S}  = 0.
 \eeq
\end{subequations}
\end{thm}
Note that the stress tensor ${\bd S}$ is not uniquely defined, although it must satisfy the equilibrium condition \rf{+14b}.  The associated density  depends only on the left stretch  of ${\bd F}$, viz., $\bd V$.  The inertial cloak corresponds to the special case of ${\bd S}= {\bd I}$, which is a trivial solution of eq. \rf{+14b}.  
The importance of Theorem \ref{thm1} is that  cloaks may simultaneously comprise  PM stiffness and  anisotropic inertia, which   provides a vastly richer potential  set of 
material parameters, not limited to the model of eq. \rf{933}.

Theorem \ref{thm1} implies that the phase speed,  wave velocity vector and polarization (not normalized)   for plane waves with phase direction ${\bd n}$ are, from eq. \rf{-104} and \rf{37}, 
\beq{-105}
v^2 =     {\bd n}\cdot  {\bd V}^2 {\bd n} , 
\qquad
{\bd c} = v^{-1} \,   {\bd V}^2 {\bd n},
\qquad
{\bd q} = {\bd S}^{-1}{\bd c} . 
\eeq
    The phase speed and wave velocity are independent of whether the cloak is an IC or the generalized PM-IC. These important wave properties are functions of the deformation only.  They can be expressed in revealing forms  using the deformation gradient as  $v = |   {\bd F}^t{\bd n}|$ and ${\bd c} = {\bd F}{\bd N}$, where ${\bd N} = {\bd F}^t{\bd n}   /|   {\bd F}^t{\bd n}|$.    Note that the  polarization ${\bd q}$ does in general  depend upon the PM properties through the stress ${\bd S}$.  
    
\subsubsection{Continuity between the cloak and the acoustic fluid}

Continuity  conditions at the cloak outer surface in the physical description follow in the same manner as \rf{538}.  The main difference is that the stress in the cloak is not isotropic, and therefore the condition that the shear tractions on the boundary vanish must be explicitly stated.    The conditions for the pseudo-pressure which satisfies eq. \rf{202} are 
\beq{554}
\big[  {\bd n}  {\bd S} p\big] = 0, \qquad 
\big[ {\bd n}\cdot {\bd \rho}^{-1} {\bd S} \nabla p  \big]=0
\quad \text{on } \partial \omega_+ .  
\eeq
These follow from eqs. \rf{-21} and \rf{1}. 

\subsubsection{Rays in the cloak are straight lines  in the undeformed space}    
Although Theorem \ref{thm1} implies that the simple wave equation \rf{+44} in  $\Omega$ is exactly mapped to eq. \rf{202} in  $\omega$ and hence all wave motion properties transform accordingly, including rays, it is instructive to deduce the ray transformation separately. 
We now demonstrate explicitly that  rays in the cloak $\omega$, which are curves that minimize travel time, are just straight lines in $\Omega$.  Consider the  straight line: ${\bd X}(t) = {\bd X}_0 + \tau {\bd N}$, where ${\bd N}$ is a  unit vector in $\Omega$. The associated curve in $\omega$
is ${\bd x}(\tau) = {\bd x} ({\bd X}_0 + \tau{\bd N})$.  Differentiation yields
$
 {\dd {\bd x}}/{\dd \tau } 
= {\bd F}  {\bd N} $, or 
$
 {\dd {\bd x}}/{\dd \tau }=   {\bd V}^2{\bd s}$
where the vector ${\bd s}$ is defined  ${\bd s} \equiv {\bd F}^{-t}  {\bd N}$. Differentiating 
${\bd s}(\tau)$, keeping in mind that ${\bd N}$ is fixed, gives
\beq{582}
\frac{\dd {\bd s}}{\dd \tau } 
=    \frac{\dd {\bd F}^{-t}}{\dd \tau } {\bd F}^t {\bd s}  
=  {\bd s}{\bd V}^2{\bd F}^{-t} \big( \frac{\dd {\bd x}}{\dd \tau }\cdot\nabla \big) {\bd F}^{-t}
= \frac12 {\bd s}{\bd V}^2  ( \nabla {\bd V}^{-2} ) \frac{\dd {\bd x}}{\dd \tau }, 
\eeq
where the compatibility identity $\partial F^{-1}_{Ij}  /\partial x_k = \partial F^{-1}_{Ik}  /\partial x_j$ has been used. We therefore deduce that  straight lines in $\Omega$ are mapped to  solutions 
 of the coupled ordinary differential equations
\beq{+574}
\frac{\dd {\bd x} }{\dd \tau} =  {\bd V}^2 {\bd s}, 
\qquad
\frac{\dd  {\bd s}}{\dd \tau} = -\frac12 {\bd s}\cdot   (\nabla{\bd V}^2 ) {\bd s} .
\eeq 
But these are identically the ray equations in the cloak, see Appendix B.  They are also the  geodesic equations for  the metric ${\bd V}^{-2}$.  The ray equations conserve the quantity ${\bd s}\cdot{\bd V}^2   {\bd s} $ which is equal to unity,  reflecting the fact that  ${\bd s}$ is the  slowness vector, ${\bd s} ={\bd n}/v$, see eqs. \rf{-105} and \rf{576}.  An illustration of rays inside the physical cloak is presented in Section \ref{sec4}. 

\subsubsection{Relation to the  Milton, Briane and Willis transformation}

 Milton, Briane and Willis \cite{Milton06} examined how the elastodynamic equations transform under general curvilinear transformations.  They  
 showed, in particular,  that if the deformation is harmonic then the   constitutive relation  \rf{931} and momentum balance \rf{932} for a compressible inviscid fluid with isotropic density transform into the equations for a pentamode material with anisotropic inertia, eqs. \rf{--1} and \rf{1}, respectively.  The deformation is harmonic if $\nabla^2_X {\bd x} = 0$, which realistically limits the transformation  to the identity \cite{Milton06}.   This would  appear to indicate acoustic cloaking using the transformation method is impossible, in contradiction to the present result.  In fact, as we show next, the MBW result is a special case of the more general theory embodied in Theorem \ref{thm1}, one that  corresponds to the choice ${\bd S}= J^{-1} {\bd V}^2 $. 
  
The PM stiffness and inertia tensor found by Milton et al. \cite{Milton06} 
are ${\bd C} = J^{-1} {\bd V}^2 \otimes {\bd V}^2$  and ${\bd \rho} = J^{-1} {\bd V}^2$ 
(their eqs. (2.12) and (2.13)).  These are of the general form required by eq. \rf{+14a}
if we identify ${\bd S}$ as ${\bd S}= J^{-1} {\bd V}^2 $.  Does this  satisfy the equilibrium condition \rf{+14b}?   Using eq. \rf{+21}, 
$\div {\bd S} = \div  J^{-1} {\bd F}{\bd F}^t = J^{-1} \Div {\bd F}^t$ and this  vanishes iff the deformation is harmonic. The MBW transformation therefore falls under the requirements of Theorem \ref{thm1}  for the specific choice of ${\bd S}= J^{-1} {\bd V}^2 $ which  satisfies the equilibrium equation \rf{+14b} only if the deformation is harmonic. 
 
Having shown that the MBW transformation result is a special case of the present theory, it is clear that the transformation as considered here is different from theirs.  
Milton et al. \cite{Milton06} demand  that all of the equations transform isomorphically, whereas the present theory requires only  that the scalar acoustic wave equation is mapped to the 
scalar wave equation for the PM, see eqs. \rf{140}.  The mapping contains an arbitrary but divergence free tensor ${\bd S}$ which defines the particular but  {\it non-unique} constitutive relation  
\rf{931} and momentum balance \rf{932}.  Consider, for instance the displacement fields 
${\bd u}_{(X)}$ and ${\bd u}$ in $\Omega$ and $\omega$, respectively.  Under the transformation of \cite{Milton06} ${\bd u}_{(X)}\rightarrow {\bd u} ({\bd x}) = {\bd F}^{-t}{\bd u}_{(X)}({\bd X})$
 (eq. (2.2) of \cite{Milton06}).   There is no analogous constraint in the present theory.   In other words, we do not require an isomorphism between the equations for all of the field variables. Instead,  the scalar wave equation for the acoustic pressure is isomorphic to the scalar equation for the pseudo-pressure of the PM.  

\subsection{Cloaks with isotropic inertia}

Theorem \ref{thm1} opens up a vast range of potential material properties.  It means  that there is no unique cloak associated with a given transformation $\Omega \rightarrow \omega$ and its deformation gradient ${\bd F}$.   We now take advantage of this non-uniqueness to 
 consider the possibility of isotropic inertia.  Equation \rf{+14a}
indicates that the  density is   isotropic if ${\bd S}$ is  proportional to ${\bd V}$.  Hence we deduce
\begin{lem}\label{lem4}
A necessary and sufficient condition that the density is isotropic, ${\bd \rho} = 
\rho {\bd I}$,  
is that there is a scalar function
$h({\bd x})$ such that 
\beq{+15}
 \div (  h {\bd V} )= 0, 
 \eeq
 in which case
 \beq{+16}
  \rho = h^2 J,  
  \qquad 
  K=J, 
  \qquad 
  {\bd S} = h  {\bd V} , 
 \eeq
 and  the Laplacian is $ \nabla_X^2 p = 
 h J  {\bd V} :   \nabla \big(h^{-1} J^{-1} {\bd V} \nabla   p \big) $ . 
\end{lem}

There is a general  circumstance for  which a solution can be found for $h$.  It takes advantage of the second order differential equality 
\beq{+18}
 \nabla_X^2 p =   ({\bd F}^t    \nabla ) \cdot   {\bd F}^t \nabla p    .  
\eeq
Although ${\bd F} $ is generally unsymmetric,  ${\bd F} ={\bd F} ^t$ in the special case that  the deformation gradient is a pure stretch with no rotation $({\bd R}={\bd I})$.   We  therefore surmise  
\begin{lem}\label{lem5}
If the deformation gradient is  a pure stretch (${\bd R}= {\bd I}$ and  hence ${\bd F} $ coincides with $ {\bd V}$) then   
the density is isotropic, 
\beq{+20}
  \rho = J^{-1}, 
  \qquad
  K=J, 
  \qquad 
  {\bd S} = J^{-1}  {\bd V}  , 
 \eeq
 and the Laplacian becomes 
$ \nabla_X^2 p =    {\bd V} :   \nabla \big( {\bd V} \nabla p \big)  $.
\end{lem}

The infinite mass problem of the IC can be avoided  if the material near the inner boundary $\partial \omega_-$ has integrable mass.  This could be achieved, for instance, by requiring that the deformation near $\partial \omega_-$ is symmetric (pure stretch). Lemma \ref{lem5}
and the scaling arguments of Subsection \ref{1.4} imply that the isotropic density scales as $\rho =$O$(\eps^{d-\beta})$, which is  integrable as long as $\beta < d+1 $ $(\alpha > 1/(d+1))$.   

\subsection{Example: the rotationally symmetric cloak }
We again consider the deformation of eq. \rf{551} for the cloak $\omega = \{{\bd x}: 0< a\le |{\bd x}| \le b\}$, and 
assume the symmetric tensor ${\bd S}$ has the form 
$
{\bd S} =  w(r) \big( {\bd I}_r + \gamma(r){\bd I}_\perp \big)$.  
Differentiation yields  
$
\div {\bd S} = \big[ w' - (d-1)(\gamma -1) \big] \hat{\bd x}$,   
and the ``equilibrium" condition \rf{+14b} is  satisfied if $w(r)$ and $\gamma(r)$ are related by 
$w' = (d-1)(\gamma -1)$.    It is  convenient to introduce a new function $g (r)$ such that 
$\gamma =  {rg' }/{ g}$ and 
$w  =  (  {g }/{r}  )^{d-1} $, which automatically makes  $\div {\bd S} =0$. 
The  cloak parameters therefore have general rotationally symmetric  form 
\beq{616}
K = \frac1{f'} \big( \frac{r}{f}\big)^{d-1} , 
\quad
{\bd S} =   \big( \frac{\g }{r} \big)^{d-1} \big[ {\bd I}_r + \frac{r\g' }{ \g} {\bd I}_\perp \big],
\quad
{\bd \rho} =   f' \big( \frac{\g^2}{rf} \big)^{d-1} \bigg[ {\bd I}_r +  
\big( \frac{ f \g'}{f' \g} \big)^2 
{\bd I}_\perp \bigg]. 
\eeq

The functions $f$ and $\g$ are independent of one another, and together define a 2-degree of freedom class of PM-IC cloaks.  The general  solution has both anisotropic stiffness and anisotropic inertia.   The previous example of the pure IC corresponds to the special case of $\g = r$, for which eq. \rf{616} gives 
${\bd S}={\bd I}$ and $K$ and ${\bd \rho}$ agree with eq. \rf{553}.

The form of the stress ${\bd S}$ indicates the PM-IC has transversely isotropic (TI) symmetry.  This is a special case of the orthotropic PM considered earlier.  
A normal TI solid  with   axis of symmetry  in the $x_3-$direction has five independent 
elastic moduli: $C_{11}$, $C_{33}$, $C_{12}$, $C_{13}$ and $C_{44}$.   The last is a shear modulus, the other shear modulus is $C_{66} = (C_{11}- C_{12} )/2$.   We set all shear moduli to zero - implying  $C_{44}= 0$ and $C_{12} = C_{11}$, and the remaining independent moduli $C_{11}$, $C_{33}$  and  $C_{13}$ satisfy 
$C_{11}C_{33}-C_{13}^2 = 0$.
The PM  therefore has two independent elastic moduli. 
 Let 
$C_{33}\rightarrow K_r (r) $,  $C_{11}\rightarrow K_\perp (r) $ and  $C_{13}\rightarrow 
\sqrt{ K_r K_\perp} $,  
then the fourth order  elasticity tensor defined by \rf{616} is 
\beq{701}
{\bd C} = K {\bd S} \otimes {\bd S} = (\sqrt{K_r} {\bd I}_r +\sqrt{K_\perp} {\bd I}_\perp) \otimes  (\sqrt{K_r} {\bd I}_r +\sqrt{K_\perp} {\bd I}_\perp), 
\eeq
where the stiffnesses $K_r$, $K_\perp$, and the principal values of the  inertia tensor  given by eq. \rf{555}, are 
\beq{10}
K_r = \frac{ 1 }{   f'} \big( \frac{\g^2}{rf}\big)^{d-1},
\quad
K_\perp   = \frac{ r {\g'}^2}{ ff' }\big( \frac{\g^2}{rf}\big)^{d-2},
\quad
\rho_r   =  f' \big( \frac{\g^2}{rf}\big)^{d-1}, 
 \quad
\rho_\perp = \frac{ f {\g'}^2}{ rf' }\big( \frac{\g^2}{rf}\big)^{d-2}.
\eeq
The phase speeds $c_r$ and $c_\perp$ in the principal directions are again given by eq. \rf{110}.   This might seem amazing at first sight, but recall that it is predicted from the general theory.   That is, the  phase and wave velocity speeds are independent of how we interpret the cloak material, as an IC or the more general PM-IC.   In the present example, it means that the phase and wave velocity are independent of $\g$.


\begin{table}		\label{tab2}
\begin{center}	
\caption{Behaviour of quantities near the vanishing point $r=a$ for the scaling
$f \propto \xi^\alpha $ as $\xi= r-a \downarrow 0$ with isotropic inertia.   }
\begin{tabular}{cccccc} \hline  
\, dim \,  & $c_r$ & $c_\perp$ &   $K_r$ & $K_\perp$ & $\rho$
  \\ \hline 
 2 &   $\xi^{1-\alpha}$ & $\xi^{  -\alpha}$ &  $\xi$ & $\xi^{ -1}$  & $\xi^{2\alpha -1}$
 \\
 3 &  \, $\xi^{1-\alpha}$\, &\, $\xi^{  -\alpha}$\, & \, $\xi^{1+\alpha}$ \,
&\,  $\xi^{\alpha -1}$\, & \,$\xi^{3\alpha-1}$ \,
 \\
\hline 		 
\end{tabular}
\end{center}
\end{table}

\subsubsection{Pure PM cloak with isotropic density}

The inertia is isotropic when $\rho_r = \rho_\perp$, which occurs if $\g (r) = f(r) $.   In that case
${\bd \rho} =\rho   {\bd I}$, and  eq. \rf{10} reduces to 
\beq{117}
\rho =  f' \big(\frac{f}{r}\big)^{d-1} ,
\quad
K_r = \frac{1}{ f'}\big(\frac{f}{r}\big)^{d-1},
\quad  
K_\perp =  f' \big(\frac{f}{r}\big)^{d-3}.
\eeq
We observe that the parameters of  eq. \rf{117} are obtained from  the IC parameters in eqs. \rf{553} and \rf{555} under the substitutions
$\{ K, \rho_r, \rho_\perp\}\rightarrow \{ 1/\rho, 1/K_r, 1/K_\perp\}$. 
Thus, the universal relation analogous to  eq. \rf{111}  is now 
\beq{013}
K_r K_\perp^{d-1} = \rho^{d-2},  
\eeq
and by analogy with eq. \rf{-3}  the three original  material parameters can be expressed using the phases speeds only, as
\beq{02}
\rho = c_r^{-1}   c_\perp^{-(d-1)} ,
\qquad
K_r = c_r   c_\perp^{-(d-1)} ,
\qquad
K_\perp = c_r^{-1}   c_\perp^{3-d} .
\eeq   
In summary, there is a one-to-one correspondence between the two sets of three material parameters for the limiting cases of the pure IC on the one hand, and the pure PM cloak on the other. 
Of course, as discussed before, the  density and stiffness cannot be simultaneously  isotropic.  The  PM-IC cloak with material properties \rf{616} includes both limiting cases when $\g=r$  and $\g=f$, respectively.   

 Table 2  summarizes the scaling of the physical quantities for  isotropic inertia, similar to the scalings in Table 1 for the pure IC.  Note that the wave speeds $c_r$, $c_\perp$ and the intermediate $(C_{13})$ modulus $\sqrt{K_rK_\perp}$ have limiting behaviour which is independent of the dimensionality,   while the density $\rho$ and the moduli $K_r$ and $K_\perp$ depend upon whether the cloak is in 2D or 3D.

\section{Further examples  }\label{sec4}

\subsection{A non-radially symmetric cloak with finite mass}
The examples considered above are rotationally symmetric and rather special in that they can be made using uniformly pure IC, or pure PM, or hybrid  PM-IC.    The pure IC model is always achievable as Lemma \ref{lem1} showed, but it suffers from the infinite mass catastrophe.   The pure PM model requires that Lemma \ref{lem4} hold at all points, which is not realistic.  However, we can always obtain a cloak comprising partly pure PM by requiring the deformation to be locally a pure stretch (Lemma \ref{lem5}).  In particular, by constraining the deformation near the inner surface  $\partial \omega_-$ in this manner, the density can be made both isotropic and integrable.   We now demonstrate this for a non-rotationally symmetric cloak.

For  ${\bd A}\in$Sym$^+$ and   $h(\zeta), h'(\zeta)>0 $ for $ \zeta \in [0,1]$, consider the deformation 
\beq{557}
{\bd x} = \zeta^{-1}h(\zeta){\bd A}{\bd X},
\qquad \zeta = ({\bd X}\cdot {\bd A}{\bd X})^{1/2}.  
\eeq
This generalizes the deformation of eq. \rf{551} $( {\bd A}= {\bd I})$ and has the important property that the deformation gradient is symmetric, 
\beq{558}
{\bd F} = \frac{h}{\zeta}{\bd A}
+ \frac{1}{\zeta} \big(\frac{h}{\zeta}\big)'\ ({\bd A}{\bd X})\otimes ({\bd A}{\bd X}). 
\eeq
  The inner surface is an ellipse (2D) or ellipsoid (3D),
\beq{093}
\partial \omega_- = \{ {\bd x}: \quad {\bd x}\cdot {\bd A}^{-1}{\bd x} =h^2(0)\}.
\eeq 

The mapping ${\bd X}\leftrightarrow {\bd x}$ must be the identity on the outer surface of the cloak $\partial \omega_+ = \partial \Omega$.   This eliminates the transformation \rf{557} as a possible deformation in the vicinity of $\partial \omega_+$ but it does not rule it out elsewhere.  In particular, it can be used on the inner surface $\partial \omega_-$  and for a finite  surrounding volume.  Then it could be patched to a different mapping closer to  the outer boundary of the cloak,  one which reduces to the identity on  $\partial\omega_+$.  For instance, 
\beq{560}
{\bd x} = \zeta^{-1}h(\zeta) {\bd A}^{\nu} {\bd X},
\eeq
where ${\nu} ({\bd x}) = 1 $ for all ${\bd x}$ between $\partial\omega_-$ and some surface ${\cal C} $, beyond which  
${\nu}$ decreases smoothly to zero as ${\bd x}$ approaches $\partial\omega_+$, which is assumed to be a level surface of $\zeta$, i.e. an ellipsoid   or an ellipse.  We assume  that  $\zeta =1$ on the outer surface, so that 
\beq{094}
\partial \omega_+ = \{ {\bd x}: \quad {\bd x}\cdot {\bd A}{\bd x} =h^2(1)\}. 
\eeq 
Let ${\cal C} $ be the level surface  $\zeta = \zeta_0$ for constant $\zeta_0\in(0,1)$.  The  
 surface separating the pure PM inner region from the PM-IC outer part of the cloak is  therefore
\beq{095}
{\cal C} = \{ {\bd x}: \quad {\bd x}\cdot {\bd A}^{-1}{\bd x} = h^2(\zeta_0) \}.
\eeq 
Based on Lemma \ref{lem5} the inner part of the cloak between $\partial\omega_-$ and ${\cal C} $ can be constructed from pure PM material with isotropic density 
$\rho = J^{-1}= (|{\bd A}| h')^{-1}  \big({\zeta}/{h}\big)^{d-1}$.  The remaining  part of the cloak is PM-IC, and the  mass of the entire cloak will be finite. 

\begin{figure*}[htbp]
\centering
\includegraphics[width=3.5in , height=3.0in 					]{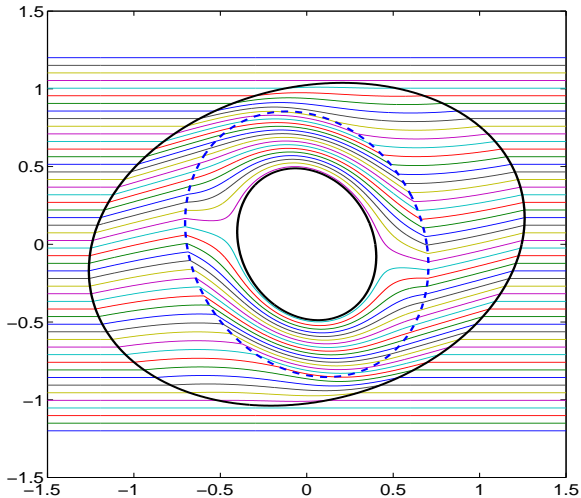} 
	\caption{Ray paths through a non-radially symmetric cloak.  The solid curves are the inner and outer surface of the cloak.  The dashed curve delineates the inner region in which the deformation gradient  is everywhere symmetric  and the  cloak is  pure PM with finite isotropic mass.   Two  movies of the rays and wavefronts in this cloak may be viewed \href{http://rci.rutgers.edu/~norris/cloaking.php}{here}.
	} 
		\label{f2}  
\end{figure*}

For instance, in Figure \ref{f2}:  $h(\zeta) = \frac12(1+\zeta)$ for $\zeta \in [0,1]$;  ${\nu} = 1$  for  $\zeta \in [0,\frac34 )$  and ${\nu} = 4(1-\zeta)$ for $\zeta \in [\frac34 ,1]$, and
the principal values of ${\bd A}$ are $0.6 $ and $1.0$.   The Figure shows 
 each ray following a continuous path through the cloak, with collinear incident and emergent ray paths. There is a unique ray separating the rays  traversing the cloak in opposite senses, and which defines a ``stagnation point" at the cloak inner surface.  The separation ray is the one that would intersect  the  singular point in the undeformed space, ${\bd O}$ in Figure \ref{f1}.   This is the origin in 
Figure \ref{f2} and since the rays are incident horizontally, the separation ray is defined by $x_2=0$ outside the cloak, and it intersects  $\partial \omega_-$ at ${\bd x} = \pm \frac12 {\bd A}{\bd e}_1 /
\sqrt{ {\bd e}_1 \cdot{\bd A}{\bd e}_1}$.   The wavefront in effect splits or tears apart  at the incident intersect and it reforms at the emergent intersect.  The time delay between these two events is infinitesimal since the tearing/rejoining is associated with the instant at which the wavefront would traverse  ${\bd O}$ in the undeformed space.  A  time-lapse movie illustrating this  may be seen \href{http://rci.rutgers.edu/~norris/cloaking.php}{here} 
(20 seconds long).  Another movie showing the ray paths for different directions of incidence can be found at the same URL.

\subsection{Scattering from near-cloaks}

A near-cloak or almost perfect cloak is defined here as one with inner surface $\partial \omega_-$ that does not correspond to the single point ${\bd X} = {\bd O}$.  We illustrate the issue using the radially symmetric deformation \rf{551}  with $f(a)$ small but non-zero, and  
 assuming time harmonic motion, with the factor $e^{-i k  t}$ understood but omitted.   Since the inner surface is not the image of a point, it is necessary to prescribe a boundary condition on the interior surface, which we take as  zero pressure on  $r=a$.   The specific nature of the boundary condition should be irrelevant as  $f(a)$ shrinks to zero. 
 
As before, the cloak occupies $\omega = \{{\bd x}:\, a\le r \le b\}$, but now $f(a)>0$.  The total response for plane wave incidence is 
$ p ({\bd x}) = p_0 e^{i k  R(r) \cos\theta }  +p_{sc}({\bd x}) $, where $p_0$ is a constant and 
\beq{41} p_{sc} =- p_0 \sum\limits_{n=0}^\infty   \begin{cases} i^n (2-\delta_{n0}) J_n( k  R(a) ) \frac{ H^{(1)}_n (k  R(r) )} { H^{(1)}_n (k  R(a) )}
\cos n \theta  ,\\ 
i^n (2n+1)  j_n( k  R(a) ) \frac{ h^{(1)}_n (k  R(r) )} { h^{(1)}_n (k  R(a) )} P_n ( \cos  \theta )  ,\end{cases}
\quad R(r) = \begin{cases} f(r), & a\le r< b, \\ 
r, & b\le r < \infty ,  \end{cases} 
\nonumber 
\eeq 
in 2D and 3D, respectively. 
A near-cloak can be defined in many ways:  for instance,  
a power law $f(r) = b\big( \frac{r-\delta}{b-\delta}\big)^\alpha$  with  $0< \delta < a$ is considered in 
\cite{Norris08c}.  Here we assume a linear near-cloak mapping similar to the one examined by Kohn et al.  \cite{Kohn08}, 
\beq{21}
f^{(\delta)}(r) = b\big( \frac{r-a}{b-a}\big) + \delta \big( \frac{b-r}{b-a}\big)
, 
\eeq
where $0<\delta < a$.  Hence,  $f^{(\delta)}(a) = \delta$ and the radius at which the mapping is zero, $r = a - \delta (b-a)/(b-\delta)$, defines the size of a smaller but  perfect cloak. 

\begin{figure}[htbp]
  \begin{center}
    \mbox{
      \subfigure[2D, $\delta = 0.3$, $\Sigma=1.48$]{\scalebox{1.0}{
      \includegraphics[width=0.45\textwidth]{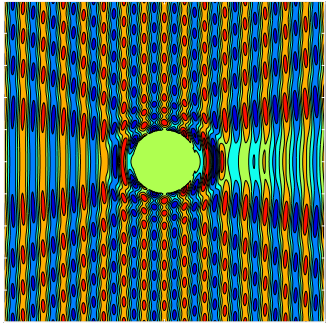}   }}
       \quad
      \subfigure[3D, $\delta = 0.3$, $\Sigma= 0.79$]{\scalebox{1.0}{ 
      \includegraphics[width=0.45\textwidth]{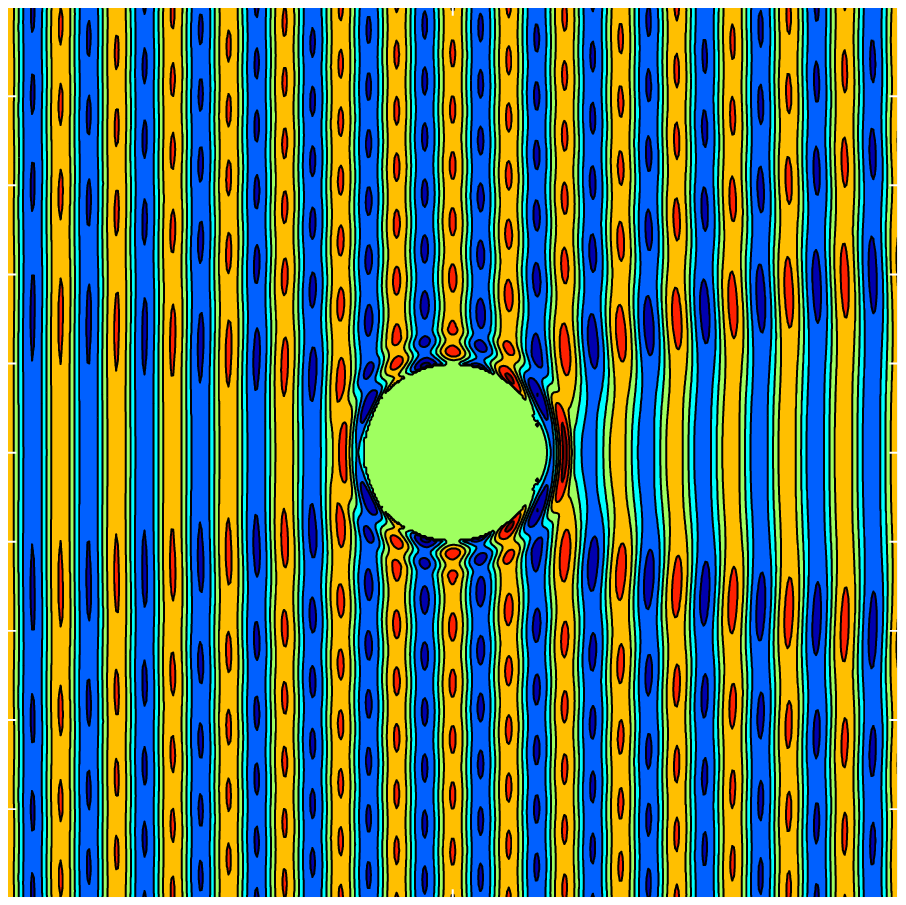}   }}
      }
      \\
    \mbox{
      \subfigure[2D, $\delta = 0.01$, $\Sigma=0.12 $]{\scalebox{1.0}{
      \includegraphics[width=0.45\textwidth]{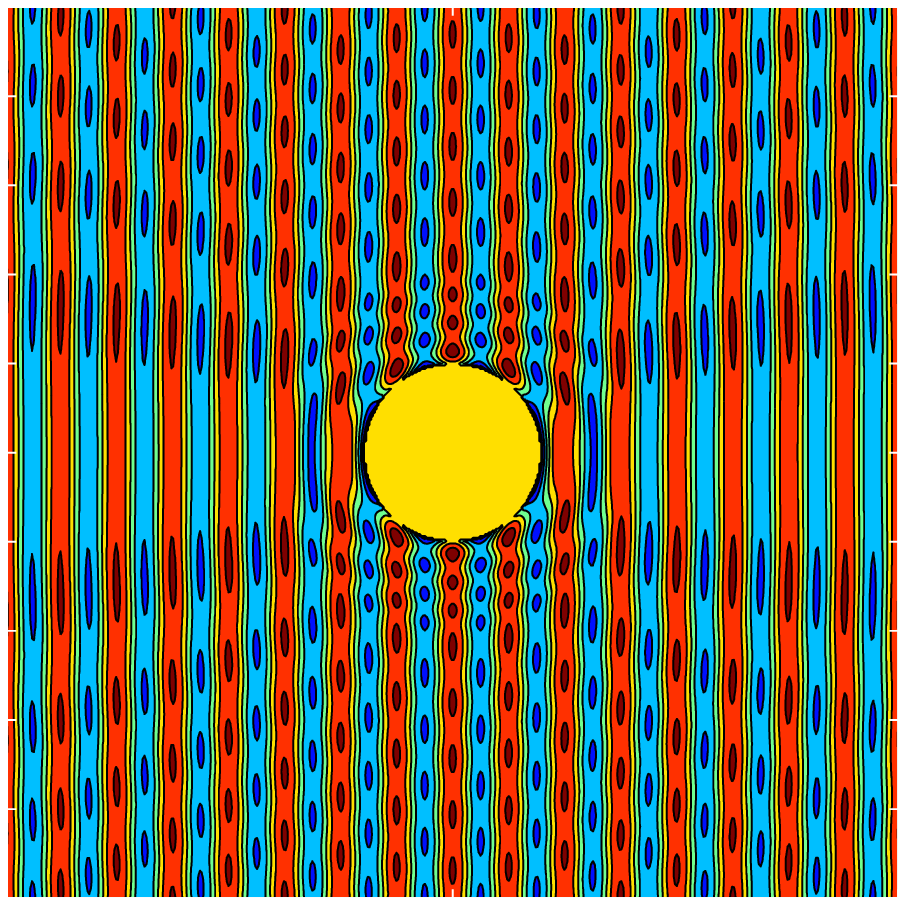}   }}      
      \quad
      \subfigure[3D, $\delta = 0.01$, $\Sigma=  1 \times 10^{-3}$]{\scalebox{1.0}{
      \includegraphics[width=0.45\textwidth]{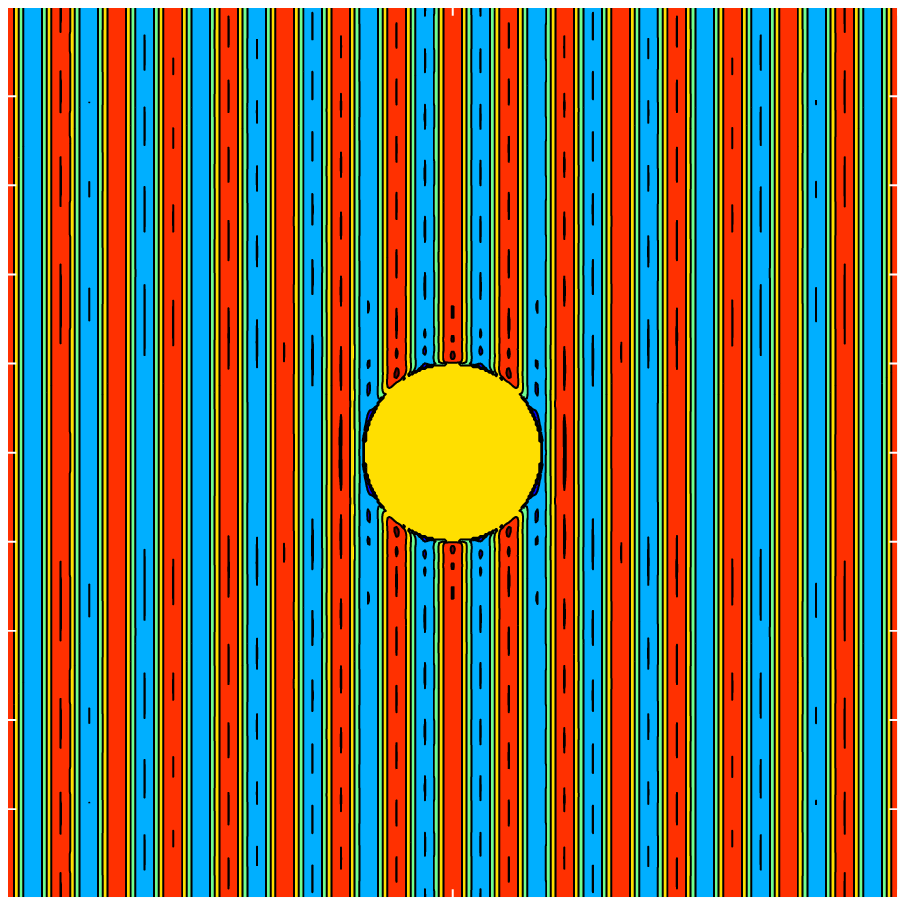}   }} 
      }
    \caption{A  plane wave is incident from the left with frequency  
$k  = 10$ on the cloak defined by eq. \rf{21} with $a=1$, $b=2^{1/(d-1)}$.  The outer cloak radius $b$ is chosen so that the geometrical cross-section of the cloak is twice the geometrical cross-section of the cloaked region in both 2D and 3D.  The circular core in the plots is  the cloaked region of radius $a$.   The  virtual inner radius  $ f(a) = \delta$ is $0.3$ or  $0.01$, and $\Sigma$ is the total scattering cross-section. }
    \label{f3}
  \end{center}
\end{figure}

  Some representative results are shown in Figure \ref{f3}, 
which  illustrates clearly a disparity between cylindrical and spherical cloaking, even when the physical optics cross-sections are identical.  Thus,  for $f(a) =0.01 a$ the 3D cross-section is negligible   (Fig.  \ref{f3}.d) but for 2D the cross-section is two orders of magnitude  larger (Fig.  \ref{f3}.c).   Ruan et al. \cite{Ruan07} found that  the perfect cylindrical EM cloak is sensitive to perturbation.  This sensitivity is  evident  from the present analysis through the dependence on the length  $\delta $ which measures the departure from perfect cloaking $(\delta =0)$.  


The ineffectiveness of the same cloak in 2D as compared with 3D can be understood in terms of the scattering cross-section. 
The leading order far-field is of the form $p = p_0 e^{i k  z} + p_0 g(\hat{\bd x}) r^{-(d-1)/2} (i 2 \pi/k )^{(3-d)/2}
e^{ik  r} $.  The optical theorem implies that the total scattering cross-section,  and hence the total energy scattered, is determined by  the forward scattering amplitude: $\Sigma = 4\pi k^{-1} \text{Im}g(\hat{\bd e}_z)$.   Thus, 
\beq{51}
\Sigma = 
\begin{cases}
\frac{ 4 }{k } \sum\limits_{n=0}^\infty  (2- \delta_{n0}) \text{Re} 
\frac{ J_n (k  f(a) )} { H^{(1)}_n (k  f(a) )}, & 2D,
\\ 
\frac{ 4\pi}{k^2} 
\sum\limits_{n=0}^\infty  (2n+1) \text{Re} 
\frac{ j_n (k  f(a) )} { h^{(1)}_n (k  f(a) )}, & 3D.
\end{cases}
\eeq
The cross-section  is dominated in the small $k  f(a)$ limit by the $n=0$ term, with leading order approximations 
\beq{52}
\Sigma = 
\begin{cases}
\frac{\pi^2}{k } |\ln k  f(a)|^{-2} + \ldots , & 2D,
\\
4\pi f^2(a) + \ldots , & 3D.
\end{cases}
\eeq
This explains the greater efficacy in 3D,  and suggests that all things being equal, cylindrical cloaking is  more difficult to achieve than its spherical counterpart. 


\section{Discussion and conclusion} \label{sec5}

Starting from the idea of an acoustic cloak defined by a finite deformation we have shown that the acoustic wave equation in the undeformed region is mapped into a variety of possible equations in the physical cloak.   Theorem \ref{thm1} implies that the  general form of the wave equation in the cloak is 
\beq{202+}
 K {\bd S} :\nabla \big(  {\bd \rho}^{-1} {\bd S} \nabla  p     \big) -  \ddot{p} =0, 
\eeq
where the stress-like symmetric tensor ${\bd S}$ is divergence free and the inertia tensor is 
${\bd \rho} = J  {\bd S} {\bd V}^{-2} {\bd S}$.   The non-unique nature of ${\bd S}$ for a given fixed deformation opens  many possibilities for interpreting the cloak in terms of material properties. 

If ${\bd S}$ is constant (${\bd S}= {\bd I}$ with no loss in generality) then the cloak material corresponds to an acoustic fluid with pressure $p$ defined by a single bulk modulus but with a mass density ${\bd \rho}$ that is  anisotropic, which we call the inertial cloak (IC).  The IC model is mathematically  consistent but physically impossible because it requires a cloak of infinite total mass. There appears to be no way to avoid this  if one restricts the cloak material properties to the IC model.  If one is willing to use an imperfect cloak with finite mass, and is concerned with fixed frequency waves, then the scattering examples 
show that  significant cloaking can be obtained by shrinking the effective visible radius to be subwavelength.   The 2D and 3D responses for imperfect cloaking are quite distinct, with far better results found in 3D. 

A cloak of finite mass is achievable by allowing ${\bd S}$ to be spatially varying and divergence free.  The general material associated with eq. \rf{202+},  called PM-IC, has both anisotropic inertia and anisotropic elastic properties.  The elastic stiffness tensor has the form of a pentamode material (PM) characterized by the symmetric tensor 
${\bd S}$ and a single modulus $K$. Under certain circumstances, characterized in Lemmas \ref{lem4} and \ref{lem5}, the density becomes isotropic and the material is pure PM.  More importantly, the total mass can be made finite.    

The finite mass problem arises from how we interpret the cloak material in the neighbourhood of its inner surface.  It is therefore not necessary to totally abandon the pure IC model, but it does mean that the alternative PM-IC is required at the inner surface.  From the examples considered here it appears that one can always use a pure PM model near the inner cloak surface, and thereby achieve finite mass.  One method is to force the deformation near the inner surface to be a pure stretch, then Lemma \ref{lem5} implies that the density is locally $\rho = 
1/\det {\bd F}$.  The total mass  remains finite as long as $\rho$ is locally  integrable, which is easily achieved.   

The theory and simulations  of PM-IC and PM materials  presented here illustrate the wealth of possible material properties that are opened up through the general PM-IC model of acoustic cloaking.  Physical implementation is in principle feasible: for instance, anisotropic inertia can be achieved by microlayers of inviscid acoustic fluid \cite{Schoenberg83}, while the microstructure required for 
pentamode  materials has been described \cite{Milton95}.   It remains to combine these known methods to obtain  practical PM-IC materials.

\subsubsection*{Acknowledgments} 
Constructive suggestions from the anonymous reviewers are appreciated. 

\appendix

\Appendix{Proof of Theorem \ref{thm1}}

A weak but instructive form of the identity \rf{+13} is proved first. 
Consider the possible identity 
\beq{404}
\nabla_X^2 p = c {\bd A} : 
 \nabla \big( {\bd B} ^{-2} \div  c^{-1} {\bd A} p  \big), 
\eeq
where ${\bd A}$, ${\bd B}\in$Sym are non-singular and $c$ is a scalar.  Let us examine  under what circumstances this identity holds.  Let $q$ be an arbitrary test function, and consider the integral
\beq{405}
I = \int_\Omega \dd V \, q \,  c {\bd A} : 
 \nabla \big( {\bd B} ^{-2} \div  c^{-1} {\bd A} p \big).
\eeq
Substituting $\dd V  = J^{-1}\dd v$,  integrating by parts and ignoring surface contributions, yields
 \beq{406}
I = - \int_\omega \dd v  \,   \big( \div    J^{-1} c {\bd A}q  \big) \cdot 
\big(  {\bd B} ^{-2} 
 \div   c^{-1} {\bd A} p \big).
\eeq
In order to guarantee the integral is self-adjoint, that is, symmetric in both $p$ and $q$, we
demand  $c = J^{1/2}$.  The self-adjoint property is made evident by writing $I$ as 
\beq{407}
I = - \int_\omega \dd v  \,   \big(  {\bd B}^{-1} \div    J^{-1/2} {\bd A}q  \big) \cdot 
\big(  {\bd B}^{-1} \div    J^{-1/2} {\bd A}p  \big).
\eeq
If \rf{404} is to be valid, then 
\beq{408}
I = - \int_\Omega \dd V  \,    \nabla_X   q    \cdot \nabla_X p. 
\eeq
Comparing these integrals and once again using $\dd v = J \dd V $, implies 
\beq{409}
\big( J^{1/2} {\bd B}^{-1 } \div    J^{-1/2} {\bd A}q  \big) \cdot 
\big( J^{1/2} {\bd B}^{-1 } \div    J^{-1/2} {\bd A}p  \big)
=  \nabla_X   q    \cdot \nabla_X p.
\eeq
The only way that these can agree for arbitrary $p$ and $q$ is if 
\beq{410}
\div    J^{-1/2} {\bd A} = 0, 
\eeq
in which case \rf{409} becomes
\beq{411}
\big(  {\bd B}^{-1}{\bd A} {\bd F}^{-t} \nabla_X q  \big) \cdot 
\big(  {\bd B}^{-1} {\bd A} {\bd F}^{-t} \nabla_X p  \big)
=  \nabla_X   q    \cdot \nabla_X p.
\eeq
Using ${\bd V}^2 = {\bd F}{\bd F}^t$, it is clear that \rf{411}
 can only be satisfied if 
\beq{412}
  {\bd B}^{-2}  = {\bd A}^{-1}  {\bd V}^2{\bd A}^{-1} .
\eeq
A weak form of 
Theorem \ref{thm1} follows by substituting ${\bd A}= J^{1/2} {\bd P}$. 
Based upon this it is a straightforward exercise to see that the identity \rf{+13} can be derived directly by brute force differentiation of the right hand side, taking into account the constraint 
\rf{+14} and  Lemma \ref{lem1}.


\Appendix{Ray equations in an acoustic cloak}
Consider a WKB type of solution for the displacement:  ${\bd u}({\bd x},t) = {\bd U} ({\bd x} , t)e^{ik\phi (  {\bd x} , t)}$.  The leading order equation for the phase  $\phi $ and amplitude ${\bd U}$ is (see eqs. \rf{-21}-\rf{-104})   
\beq{571}
\big[ K\,  ({\bd S}{\nabla \phi })\otimes ({\bd S}{\nabla \phi}) - \dot{\phi}^2 {\bd \rho}
\big]  {\bd U} =0. 
\eeq
The inner product of eq. \rf{571} with ${\bd U} $ may  be written 
$H_+H_- = 0$ where
\beq{572}
H_\pm = \dot{\phi} \pm K^{1/2} | \hat{\bd q}\cdot {\bd \rho}^{-1/2}{\bd S}{\nabla \phi}|,  
\eeq
and  $\hat{\bd q} = {\bd \rho}^{1/2}{\bd U}/ |{\bd \rho}^{1/2}{\bd U}|$.
We focus on the characteristic $H_+ ( \dot{\phi}, \nabla \phi, {\bd x})= 0$.  The Hamilton-Jacobi equations for this ``Hamiltonian" yield the ray equations, 
\beq{573}
\frac{\dd t}{\dd \tau} =\frac{\partial H_+}{\partial \dot{\phi} } , 
\qquad
\frac{\dd {\bd x} }{\dd \tau} =  \frac{\partial H_+}{\partial \nabla \phi}, 
\qquad
\frac{\dd \nabla \phi}{\dd \tau} = -\frac{\partial H_+}{\partial {\bd x}}, 
\qquad
\frac{\dd \dot{\phi}}{\dd \tau} = -\frac{\partial H_+}{\partial t} ,  
\eeq 
where $\tau$ is the time-like ray parameter.  Since $ {\partial H_+}/{\partial \dot{\phi} }= 1$, 
$\tau$
  may be  replaced by $t$ as the natural ray parameter, while ${\partial H_+}/{\partial t} =0$ implies that $\dot{\phi}$ is constant along a ray.  We choose  $\dot{\phi}= -1$ for convenience. 
     Define the slowness vector
${\bd s} (\tau) = \nabla \phi$ along the ray ${\bd x} = {\bd x}(\tau)$. 
The vector equation \rf{571} then implies that 
$
\hat{\bd q} = K^{1/2} {\bd \rho}^{-1/2} {\bd S} {\bd s}$,
and since $\hat{\bd q}$  is by definition a unit vector, using eq. \rf{+14a} we deduce that the slowness satisfies
\beq{576}
{\bd s}\cdot {\bd V}^2{\bd s} = 1. 
\eeq
This is simply the ellipsoidal slowness surface mentioned in Section \ref{sec3}.
Finally,  the  evolution equations along the ray can be expressed as a closed 
system for $  {\bd x}(t), {\bd s}(t)$ by using eq. \rf{573} and noting that 
$H_+ =  \big( {\bd s}\cdot {\bd V}^2{\bd s}\big)^{1/2} -1$,  
\beq{574}
\frac{\dd {\bd x} }{\dd t} =  {\bd V}^2 {\bd s}, 
\qquad
\frac{\dd  {\bd s}}{\dd t} = -  ({\bd V} {\bd s}) \cdot (\nabla {\bd V}) {\bd s}.
\eeq


\begin{thebibliography}{10}

\bibitem{Ward96}
A.~J. Ward and J.~B. Pendry.
\newblock Refraction and geometry in {M}axwell's equations.
\newblock {\em J. Modern Optics}, 43(4):773--793, 1996.

\bibitem{Post62}
E.~J. Post.
\newblock {\em Formal Structure of Electromagnetics: General Covariance and
  Electromagnetics}.
\newblock Interscience, New York, 1962.

\bibitem{Greenleaf03}
A.~Greenleaf, M.~Lassas, and G.~Uhlmann.
\newblock On nonuniqueness for {C}alderon's inverse problem.
\newblock {\em Math. Res. Lett.}, 10:685--693, Jul 2003.

\bibitem{Greenleaf03b}
A.~Greenleaf, M.~Lassas, and G.~Uhlmann.
\newblock Anisotropic conductivities that cannot be detected by {EIT}.
\newblock {\em Physiol. Meas.}, 24(2):413--419, May 2003.

\bibitem{Pendry06}
J.~B. Pendry, D.~Schurig, and D.~R. Smith.
\newblock Controlling electromagnetic fields.
\newblock {\em Science}, 312(5781):1780--1782, June 2006.

\bibitem{Leonhardt06}
U.~Leonhardt.
\newblock Optical conformal mapping.
\newblock {\em Science}, 312(5781):1777--1780, June 2006.

\bibitem{Cummer07}
S.~A. Cummer and D.~Schurig.
\newblock One path to acoustic cloaking.
\newblock {\em New J. Phys.}, 9(3):45+, March 2007.

\bibitem{Chen07}
H.~Chen and C.~T. Chan.
\newblock Acoustic cloaking in three dimensions using acoustic metamaterials.
\newblock {\em Appl. Phys. Lett.}, 91(18):183518+, 2007.

\bibitem{Cummer08}
S.~A. Cummer, B.~I. Popa, D.~Schurig, D.~R. Smith, J.~Pendry, M.~Rahm, and
  A.~Starr.
\newblock Scattering theory derivation of a {3D} acoustic cloaking shell.
\newblock {\em Phys. Rev. Lett.}, 100(2):024301+, 2008.

\bibitem{Milton06}
G.~W. Milton, M.~Briane, and J.~R. Willis.
\newblock {On cloaking for elasticity and physical equations with a
  transformation invariant form}.
\newblock {\em New J. Phys.}, 8:248--267, 2006.

\bibitem{Schoenberg83}
M.~Schoenberg and P.~N. Sen.
\newblock Properties of a periodically stratified acoustic half-space and its
  relation to a {B}iot fluid.
\newblock {\em J. Acoust. Soc. Am.}, 73(1):61--67, 1983.

\bibitem{Sheng07}
J.~Mei, Z.~Liu, W.~Wen, and P.~Sheng.
\newblock Effective dynamic mass density of composites.
\newblock {\em Phys. Rev. B}, 76(13):134205+, 2007.

\bibitem{Torrent08}
D.~Torrent and J.~S\'{a}nchez-Dehesa.
\newblock Anisotropic mass density by two-dimensional acoustic metamaterials.
\newblock {\em New J. Phys.}, 10(2):023004+, 2008.

\bibitem{Milton07}
G.~W. Milton and J.~R. Willis.
\newblock On modifications of {N}ewton's second law and linear continuum
  elastodynamics.
\newblock {\em Proc. R. Soc. A}, 463(2079):855--880, March 2007.

\bibitem{Schurig06}
D.~Schurig, J.~J. Mock, B.~J. Justice, S.~A. Cummer, J.~B. Pendry, A.~F. Starr,
  and D.~R. Smith.
\newblock Metamaterial electromagnetic cloak at microwave frequencies.
\newblock {\em Science}, 314(5801):977--980, November 2006.

\bibitem{Greenleaf07}
A.~Greenleaf, Y.~Kurylev, M.~Lassas, and G.~Uhlmann.
\newblock Full-wave invisibility of active devices at all frequencies.
\newblock {\em Comm. Math. Phys.}, 275(3):749--789, November 2007.

\bibitem{Greenleaf08}
A.~Greenleaf, Y.~Kurylev, M.~Lassas, and G.~Uhlmann.
\newblock {Comment on "{S}cattering theory derivation of a 3{D} acoustic
  cloaking shell"}.
\newblock  Jan 2008.  \href{http://arxiv.org/abs/0801.3279v1}{URL}

\bibitem{Milton95}
G.~W. Milton and A.~V. Cherkaev.
\newblock Which elasticity tensors are realizable?
\newblock {\em J. Eng. Mater. Tech.}, 117(4):483--493, 1995.

\bibitem{Ogden84}
R.~W. Ogden.
\newblock {\em Non-Linear Elastic Deformations}.
\newblock Dover Publications, 1997.

\bibitem{Kohn08}
R.~V. Kohn, H.~Shen, M.~S. Vogelius, and M.~I. Weinstein.
\newblock Cloaking via change of variables in electric impedance tomography.
\newblock {\em Inverse Problems}, 24(1):015016+, 2008.

\bibitem{Musgrave}
M.~J.~P. Musgrave.
\newblock {\em Crystal Acoustics}.
\newblock Acoustical Society of America, New York, 2003.

\bibitem{Norris08c}
A.~N. Norris.
\newblock Acoustic cloaking in 2{D} and 3{D} using finite mass,
\newblock  Feb 2008.  \href{http://arxiv.org/abs/0802.0701}{URL}

\bibitem{Ruan07}
Z.~Ruan, M.~Yan, C.~W. Neff, and M.~Qiu.
\newblock Ideal cylindrical cloak: Perfect but sensitive to tiny perturbations.
\newblock {\em Phys. Rev. Lett.}, 99(11):113903 +, 2007.

\end{thebibliography}


\label{lastpage}
\end{document}